\documentclass[12pt,a4paper,english,nofootinbib,showkeys,superscriptaddress,aps]{revtex4}
\usepackage[english]{babel}
\usepackage{graphicx,color}
\usepackage{latexsym}
\usepackage{amsmath}
\usepackage{amsfonts}
\usepackage{amssymb,slashed}
\usepackage{bbold}
\usepackage[latin1]{inputenc}
\usepackage{verbatim}
\usepackage{hyperref}

\DeclareMathOperator{\Tr}{Tr}

\def\be{\begin{equation}}
\def\ee{\end{equation}}
\def\beq{\begin{eqnarray}}
\def\eeq{\end{eqnarray}}

\begin{document}

\title{\Large Entropy from Scaling Symmetry Breaking}

\author{Neymar Cavalcante}
\email{neymarnepomuceno@gmail.com}

\affiliation{Instituto de F\'{\i}sica, Universidade de Bras\'{\i}lia, Caixa Postal 04455, 70919-970, Bras\'{\i}lia, DF,Brazil}

\author{Saulo Diles}

\email{smdiles@gmail.com}

\affiliation{Instituto de F\'{\i}sica - Universidade Federal do Rio de Janeiro, Caixa Postal 68528, 21945, Rio de Janeiro, RJ, Brazil}

\author{Kumar S. Gupta} 
\email{kumars.gupta@saha.ac.in}

\affiliation{Theory Division, Saha Institute of Nuclear Physics, 1/AF Bidhannagar, Kolkata 700064, India}

\author{Amilcar R. de Queiroz}
\email{amilcarq@unb.br}  

\affiliation{Instituto de F\'{\i}sica, Universidade de Bras\'{\i}lia, Caixa Postal 04455, 70919-970, Bras\'{\i}lia, DF,Brazil}

\affiliation{Departamento de F\'{\i}isica Te\'orica, Facultad de Ciencias, Universidad de Zaragoza, 50009
Zaragoza, Spain}


\begin{abstract}
The scaling symmetry in conformal quantum mechanics (CQM) can be broken due to the boundary conditions that follow from the requirement of a unitary time evolution of the Hamiltonian.
We show that the scaling symmetry of CQM can be restored by introducing a suitable mixed state, which is associated with a nonvanishing von Neumann entropy. We give an explicit formula for the entropy arising from the mixed state in CQM. Our work provides a direct link between the restoration of a broken symmetry and von Neumann entropy.
\end{abstract}


\maketitle

\section{Introduction}
\label{sec:1}

The representations of $so(2,1)$ Lie algebra describe the physical states of a system governed by conformal quantum mechanics (CQM) \cite{fubini}. In such systems, the Hamiltonian $H$ together with the generators of dilatation $G$ and the special conformal transformation $K$ span the spectrum generating algebra. 
The $so(2,1)$ Lie algebra appears in the description of a large class of physical systems, including molecules \cite{PhysRevLett.87.220402,PhysRevD.68.125013,pulak}, black holes \cite{Birmingham2001a,ksg1,Chakrabarti2008},  
graphene \cite{graph1,graph2,graph3} and various types of rational Calogero models \cite{calo1,calo2,calo3}. It also appears in the study of instabilities of Coulomb phase in QCD and confinement \cite{Asorey:2012hqa,Asorey:2014kma,Asorey:2014uaa}. It is also associated with renormalization group \cite{Gupta1993} and dimensional transmutation \cite{jackiw1991delta,camblong2001dimensional,CamblongAnn.Phys.287:57-1002001} in quantum mechanics. It is a remarkable fact that the essential physical features of such a large class of apparently unrelated physical systems are characterized by the representations of $so(2,1)$ Lie algebra.

The Hamiltonian in CQM is an unbounded operator on a Hilbert space, which requires a specification of its domain \cite{reed}. The domain, or equivalently the boundary conditions, are obtained by demanding a unitary time evolution, which is generated by a self-adjoint Hamiltonian.  The boundary conditions leading to a unitary time evolution in CQM may not be unique. For certain values of the system parameters, the Hamiltonian of a CQM may admit a family of self-adjoint extensions \cite{calo1}. The corresponding domains are labelled by a self-adjoint extension parameters \cite{reed}. In that case, the action of the dilatation in general would not preserve the domain of the Hamiltonian, leading to a quantum mechanical symmetry breaking \cite{calo1}. In other words,  scaling symmetry cannot be implemented on a unitary irreducible representation (UIR), labelled by a given domain, or equivalently by a given self-adjoint extension of the Hamiltonian. The action of the dilatation will map one self-adjoint extension parameter to another, thereby indicating the breakdown of the scaling symmetry \cite{PhysRevD.66.125013}. Such a symmetry breaking happens purely due to quantum effects \cite{PhysRevD.34.674}.

Given this scenario, it is natural to ask if there exists a possibility of a restoration of the broken symmetry, within the ambit of CQM. In this paper we demonstrate that the restoration of such a broken symmetry due to the self-adjoint extension of $H$ is possible at the expense of the introduction of a mixed state. The mixed state is compatible with the unitary time evolution. However, in general it will lead to a nonvanishing von Neumann entropy. Thus, in our picture, the restoration of the broken scaling symmetry in CQM is directly related to the generation of von Neumann entropy. For discrete symmetries similar ideas were discussed before in \cite{Balachandran2012,Gupta2013a}. One of the new features of this work is that we show how to restore a continuous scaling symmetry through the introduction of a mixed state. We also obtain an entropy formula similar to that of Cardy \cite{cardy} for the case of conformal systems with  Virasoro algebra with a central charge \cite{Birmingham2001a,ksg1,Chakrabarti2008}. Cardy formula uses Virasoro algebra with a central charge and modular invariance \cite{cardy}. We rather use the idea of mixed states to derive the entropy formula. One advantage of our formulation is that it simply uses the $so(2,1)$ Lie algebra structure. This makes our approach amenable to a wider class of applications independent of the Virasoro algebra. 

Recently there have been empirical studies connecting symmetry breaking to entropy \cite{roldan-nature}. It is plausible that our work could be related to such experiments.

The paper is organized as follows. In section 2, we present the CQM of a massless particle on a plane. We then discuss the appearance of a non-trivial family of boundary conditions that renders the Hamiltonian as an appropriate self-adjoint operator. In section 3, we construct the mixed state that restores the scaling symmetry. In section 4, we compute the entropy of the mixed state that has been used to restore the broken scaling symmetry. This computation consists basically in the appropriate definition of the measure in the space of boundary conditions over which one averages pure states to restore the anomalous symmetry. Section 5 concludes the paper with some discussions and an outlook.

\section{Particle on a plane with a defect : a Model for CQM}

A system is described by CQM if the corresponding Hamiltonian $H$ together with the generator $G$ of dilatations and the generator $K$ of special conformal transformations satisfy the $so(2,1)$ Lie algebra with commutators 
\begin{align}
	    \label{so21-alg-1}
            i \left [ G , H \right ] &= H, \\
            i \left [ K , G \right ] &= K, \\
            i \left [ H , K \right ] &= -2G.
      \end{align}
As mentioned before, a wide variety of physical systems admit such an algebraic structure. We here consider a prototype physical system that captures the essential physics of CQM. 

Our system is described by a particle moving on a plane and interacting with a point defect \cite{PhysRevD.66.125013}. Such defect consists of a flux tube perpendicular to the plane or any physical object which produces an inverse square-interaction. The Hamiltonian is given by
\beq
\label{fullop}
H(\alpha) &=& -\frac{\partial^2}{\partial r^2} - \frac{1}{r}\frac{\partial}{\partial r} + \frac{\alpha}{r^2} - \frac{1}{r^2}\frac{\partial^2}{\partial \varphi^2},
\eeq 
where $(r,\phi)$ denote the polar coordinates on the plane and the dimensionless parameter $\alpha$ captures the strength of the inverse-square potential. We redefine the variables, so that this Hamiltonian is effectively written as 
\beq
\label{fullop-2}
H(\alpha) &=& -\frac{\partial^2}{\partial r^2} - \frac{1-4\alpha}{4r^2} - \frac{1}{r^2}\frac{\partial^2}{\partial \varphi^2}.
\eeq 
We have basically transformed the initial measure $r^2 dr$ for the radial part to just $dr$. Henceforth we will consider only the latter.

We next consider the eigenvalue equation is $H(\alpha)\psi=E(\alpha) \psi$ with the the separation of variables $\psi (r, \phi) = f (r) ~ \chi (\phi)$. The effective angular Hamiltonian 
\be
\label{angular-op}
H_\phi = - \frac{1}{r^2}\frac{\partial^2}{\partial \varphi^2}
\ee
is self-adjoint for the domains
\be
\label{ang-domain}
D_\theta=\left\{ L^2(S^1,d\varphi):~\chi(2\pi)=e^{i\theta} \chi(0), ~ \chi'(2\pi)=e^{i\theta} \chi'(0)\right\}, 
\ee
parametrized by $\theta\in [0,2\pi)$. The spectrum of this angular operator is $\lambda_n=(n+\frac{\theta}{2\pi})^2$,  $n\in\mathbb{Z}$. This provides a one-parameter family of inequivalent quantization of the angular operator in (\ref{angular-op}).

As discussed in \cite{Balachandran2012,Gupta2013a}, the self-adjoint extension parameter $\theta$ breaks the parity $P$ and time reversal $T$ symmetries, except when $\theta = 0, \pi$. It was also proposed in \cite{Balachandran2012} that an appropriate impure (or mixed) state can be used to restore the $P$ and $T$ symmetries and the associated RG procedure was discussed in \cite{Gupta2013a}. Here we shall consider the radial dynamics in detail and see that that the boundary conditions can break the continuous scaling symmetry of the system.


The eigenvalue problem for the radial Hamiltonian is 
\be
\label{radial-eqn}
H_r g(r)  \equiv \left [ -\frac{\partial^2}{\partial r^2} + \frac{\delta}{r^2} \right ] g(r) = E g(r),
\ee
where
\be
\label{gamma}
\delta = - \frac{1}{4} + \lambda_n + \alpha.
\ee
The radial operator $H_r$ is symmetric in the domain $D(H_r) \equiv \{\phi (0) = \phi^{\prime} (0) = 0,~
\phi,~ \phi^{\prime}$  absolutely continuous, $\phi \in {\rm L}^2(dr)\} $. According to the value of the parameter $\delta$, it is well-known \cite{reed} that: 
\begin{enumerate}
\item  $\delta \geq \frac{3}{4}$ : in this case $H_r$ is essentially self-adjoint in $D(H_r)$.
\item  $ -\frac{1}{4} \leq \delta < \frac{3}{4}$. In this case $H_r$ is not self-adjoint in $D(H_r)$ but admits a one-parameter family of self-adjoint extensions. The self-adjoint extensions are characterized by a real parameter $\gamma \in [0, 1]$.
\item $\delta < -\frac{1}{4}$. In this case, the ground state energy is unbounded from below and hence the system is unphysical. RG techniques can be used to address this case \cite{Gupta1993}.
\end{enumerate}

In the case where $ -\frac{1}{4} \leq \delta < \frac{3}{4}$, the boundary conditions arising from the self-adjoint extension breaks the scaling symmetry for generic values of the self-adjoint extension parameter $\gamma$ \cite{calo1}. This happens since the domain in which $H_r$ is self-adjoint is not kept invariant by the generator of dilatations. To see this, consider the action of the dilatation operator $G = \frac{-i}{2} (r \frac{d}{dr} + \frac{d}{dr} r)$
on a generic element $\phi(r) = \phi_{+}(r) + e^{i \gamma} \phi_{-}(r)
\in  \mathcal{D}_\gamma( H_r)$, where $\phi_+(r)=\sqrt{r} H^{(1)}_\nu(re^{i\pi/4})$ and $\phi_-(r)=\sqrt{r} H^{(2)}(re^{-i\pi/4})$, with $H^{(k)}_\nu$ being Hankel functions \cite{abr} and $\nu^2 = \gamma + \frac{1}{4}$. In the limit $r \rightarrow 0$, one obtains \cite{calo1}
\begin{equation*}
G \phi(r) \rightarrow
\frac{1}{{\rm sin}\nu \pi}
\left [ (1 + \nu) \frac{r^{\nu + \frac{1}{2}}}{2^\nu} \frac{ ( e^{ - i    
\frac{3 \nu \pi}{4}} - e^{i (\gamma +  \frac{3 \nu \pi}{4})})}{\Gamma (1 + \nu)}
+ (1 - \nu) \frac{r^{- \nu + \frac{1}{2}}}{2^{- \nu}}
\frac{ ( e^{ i
( \gamma +  \frac{\nu \pi}{4} ) } - e^{-i \frac{ \nu \pi}{4}}) }{\Gamma (1 -
\nu)}    
\right ].
\end{equation*}   
Now, $G \phi(r) \in \mathcal{D}_\gamma( H_r)$ only if $G\phi(r) \sim C \phi(r)$ where $C$ is a constant. This is not possible, so that $G\phi(r)$ in general does not belong to $\mathcal{D}_\gamma(H_r)$. This shows that the scaling symmetry broken at the quantum level for generic values of the self-adjoint extension parameter $\gamma$.

\section{Mixed States for CQM}

For some modes, particularly the zero mode $n=0$ when $\theta=0$ and $n=0,1$ when $\theta>0$, the dilatation operator $G$ changes the boundary conditions of the radial Hamiltonian $H_r$. Therefore, it does not make sense to implement the commutator of $G$ and $H_r$ on physical states. 

We here use a mechanism to restore broken symmetries by mixed states as suggested by Balachandran and Queiroz in \cite{Balachandran2012}, see also \cite{Balachandran:2011gj,Balachandran2012h,Queiroz2012}. Our strategy is to construct an appropriate mixed state on which the commutators between $H_r(\gamma)\equiv (H_r,\mathcal{D}_\gamma)$ and $G$ can be implemented unambiguously. Such an appropriate mixed state consists of the average over all possible boundary conditions obtained from the action of the operator $G$. 

We consider the notation $H_r(\gamma)$, with $0\leq \gamma\leq 1$. For a fixed $\mathcal{D}_\gamma$, we take a fixed general state $|\cdot;\gamma\rangle$. The mixed state that restores the scaling symmetry is of the form
\begin{equation}
	\label{MS-anomaly-restoring-1}
  \omega=\int d\mu(\gamma) ~ |\cdot;\gamma\rangle\langle \cdot;\gamma|\equiv \int d\mu(\gamma) ~ \hat{\omega}_\gamma,
\end{equation}
where $\hat{\omega}_\gamma\equiv |\cdot;\gamma\rangle\langle\cdot;\gamma|$ represents a pure state ($\hat{\omega}_\gamma^2=\hat{\omega}_\gamma$) on domain $\mathcal{D}_\gamma$ and $d\mu(\gamma)$ is a normalized measure invariant with respect to $G$. The expectation value of any observable $\mathcal{O}$ on state $\omega$ is given by
\begin{equation}
  \langle \mathcal{O} \rangle_\omega=\Tr~\omega \mathcal{O},
\end{equation}
where $\Tr$ is over the Hilbert space. If $\mathcal{O}$ is an observable invariant under $G$-action, its expectation value on state $\omega$ necessarily vanishes. In particular, by construction $[G,\omega]=0$, so that   $\langle [H_r(\gamma),G]\rangle_\omega=0$.

Another possibility of implement such broken scaling symmetry is to use projective representation \cite{Bargmann1947}. The advantage of the projective representation lies in its direct topological meaning that can be cast in terms of its non-trivial cocycle. The advantage of the mixed state lies in the simplicity of the computation of its von Neumann entropy. The explicit relation between central extensions, or projective representation, and mixed states restoring anomalies will be reported elsewhere. From the point of view of GNS construction, the appearance of mixed state associated with anomaly was also discussed in \cite{Balachandran2013g}.

\section{Entropy Computation}

Let us consider a general mixed state of the form
\begin{equation}
	\label{decomposition-3}
  \omega=\int d\mu(\gamma)~\lambda_\gamma ~ \hat{\omega}_\gamma,
\end{equation}
where $d\mu(\gamma)$ is a $U(1)$-invariant Haar measure, with $U(1)$ being isomorphic to the space of all self-adjoint extensions we are considering,  $\hat{\omega}_\gamma$ is a pure state ($\hat{\omega}_\gamma^2=\hat{\omega}_\gamma$) in the domain $\mathcal{D}_\gamma$ and $\lambda_\gamma$ is a weight satisfying
\begin{equation}
  \lambda_\gamma \geq 0, \qquad \int d\mu(\gamma)~\lambda_\gamma = 1.
\end{equation}
The von Neumann entropy of $\omega$ is given by
\begin{equation}
  S(\omega)=-\int d\mu(\gamma)~\lambda_\gamma\log \lambda_\gamma.
\end{equation}

In the previous section we argued that the requirement of covariance of $\omega$ under the action by $G$ forces the decomposition (\ref{decomposition-3}) to be uniform, that is, $\lambda_\gamma=1/\mathcal{N}$, for any $\gamma$, where $\mathcal{N}$ is an appropriate normalization factor. The proposed anomaly restoring mixed state is therefore (\ref{MS-anomaly-restoring-1}). The normalization factor is associated with the volume of a $G$-orbit in the $U(1)$-family of domains. But there are ambiguities in the definition of the entropy to be dealt with. The uniform decomposition of the mixed state $\omega$ has associated entropy
\begin{equation}
  S(\omega)= \log \mathcal{N}.
\end{equation}
Our problem now is to find $\mathcal{N}$.

Naively, $\mathcal{N}$ would be the volume of the $U(1)$-family of self-adjoint domains of the Hamiltonian. This would give $1$ if $U(1)$ is modelled after an interval with length $1$. However, the uniformity of the weights in the mixed states brings in with an ambiguity in the evaluation of $\mathcal{N}$ (a counting ambiguity). This ambiguity is resolved if one uses an appropriate Gibbs-like factor associated with the volume of the symmetry group generating this ambiguity. 

We start by considering a finite interval as a model for the $U(1)$-family of boundary conditions, that is, $0\leq \gamma \leq 1$. We proceed by considering a uniform discretization of this interval. Each discrete point of the interval is labelled by $j$ and the inter-site distance is denoted by $a$. Other regularizations of the interval leads to the same final conclusion. We keep with this uniform regularization. The regularized mixed state becomes 
\begin{equation}
 \label{mixed-state-regularized}
  \omega_N=\frac{1}{\mathcal{N}_N}\sum_{j=1}^{N} \hat{\omega}_j,
\end{equation}
where $N$ is the number of discrete points uniformly filling the interval. The discrete points $j$ can be reshuffled without affecting the counting. This implies an ambiguity in the counting associated with the symmetric group $S_N$. We then fix a choice of ordering of $j$, that is, we fix a particular $S_N$ gauge orbit. In other words, we account for the distinct manners of ordering the points $j$. The result is that $\mathcal{N}_N$ is proportional to the number of conjugacy classes of the group $S_N$. This is similar to account for the volume of orbit space of a gauge group and the fact that this volume is equal to the number of conjugacy classes of $S_N$. Recall that the number of conjugacy classes of $S_N$ is equal to the number of inequivalent irreducible representations of $S_N$. Hence, the normalization factor is $\mathcal{N}_N=(a N) |\mathcal{C}_{S_N}|$, where $|\mathcal{C}_{S_N}|$ is the number of conjugacy classes of the symmetric group $S_N$. It is a well-known result \cite{Fulton-Harris} that $|\mathcal{C}_{S_N}|=p(N)$, where $p(N)$ is the number of partitions of the integer $N$. In other words, each inequivalent irreducible representation of $S_N$ is labelled by a partition of $N$. 

The outcome of the above argument is 
\begin{equation}
	\label{regularized-entropy-1}
  S(\omega_N)=\log \mathcal{N}_N = \log (a N) p(N).
\end{equation}
We now renormalize this counting formula (\ref{regularized-entropy-1}) by taking $N\to\infty$ in a continuous manner, that is, simultaneously sending the inter-site distance $a$ to zero while keeping the unit length of the interval fixed. Therefore $a N \to 1$. At the same time, this large $N$ limit leads to the notorious asymptotic formula by Hardy-Ramanujan \cite{hardy} or Rademacher \cite{rademacher},
\begin{equation}
  p(N)\sim \frac{1}{4\sqrt{3} N}~e^{\pi\sqrt{\frac{2N}{3}}}.
\end{equation}
The entropy is therefore
\begin{equation}
  S(\omega_{N\to\infty})\sim~\pi\sqrt{\frac{2N}{3}}-\log(4\sqrt{3}).
\end{equation}
In order to compensate for the divergence in $N\to \infty$, we properly normalize the entropy by $\sqrt{N}$, so that
\begin{equation}
	\label{entropy-formula-c-equal-1}
  S_{\sqrt{N}}(\omega)=\lim_{N\to \infty}\frac{S(\omega_{N})}{\sqrt{N}}=\pi\sqrt{\frac{2}{3}}.
\end{equation}
This expression is similar to Cardy's formula for CFT with $c=1$ \cite{cardy}. We will discuss more on the relation of our derivation and the derivation of Cardy's formula in the next section.

Similar arguments work for other regularizations. The final result depends on the uniform continuous $S_\infty$ sum of pure state (\ref{MS-anomaly-restoring-1}). 

In \cite{Balachandran2013b,Balachandran2013d}, the origin of the ambiguity for the computation of an entropy of certain mixed states was extensively analysed from an algebraic perspective. It was shown there that degeneracy leads to an ambiguity of the irreducible representation of the algebra of observable. We were inspired by the above works to properly count states taking into account the mentioned ambiguity. 

\subsection{Many Particles}

The generalization of the previous counting argument to more particles is straightforward. We provide here the detailed argument.

We first consider independent distinguishable  particles. Independence means the particles do not interact among themselves. Obviously, each distinct particle independently obey conformal mechanics.

Suppose there are $c$ of such independent distinguishable particles. Then in the previous counting we have to replace $N$ to $cN$. This means that each particle carries a $U(1)$-family of self-adjoint extensions. They are independent so that we end up with $c$ of such families. Independence here leads to the fixing the form of self-adjoint family. In principle, for $c$ particles, we could have an $U(c)$ family, suggesting existence of boundary conditions that transmute one particle into another. This more general case of $U(c)$ is also associate with quantum topological change \cite{Bal-topchange, Bal-topchange2}. The new formulae become
\begin{align}
   S(\omega^c_{N}) &=\log \mathcal{N}^c_{N} = \log \left(c (Na) p(cN) \right), \\
   p(cN) &\sim \frac{1}{4\sqrt{3} cN}~e^{\pi\sqrt{\frac{2cN}{3}}}, \\
   S(\omega^c_{N\to\infty})&\sim~\pi\sqrt{\frac{2cN}{3}}-\log(4\sqrt{3}).
\end{align}
We then quotient out this last expression by $\sqrt{N}$ and take the limit $N\to \infty$, so that
\begin{equation}
	\label{entropy-formula-with-c}
  S_{\sqrt{N}}(\omega)=\frac{S(\omega_{N\to\infty})}{\sqrt{N}}=\pi\sqrt{\frac{2c}{3}}.
\end{equation}
This formula resemble that Cardy's formula for general $c$.

If the particles are indistinguishable, then we need to take extra care due to the statistics of such particles. Once the statistic group, which is associated with the fundamental group of the underlying configuration space of the system, is taken into account, the procedure of counting is the same as stated above.

\section{Conclusion}

The continuous scaling symmetry in CQM can be broken due to quantization. This happens when the boundary conditions, or equivalently the domain of the Hamiltonian is not preserved by the action of the dilatation operator \cite{PhysRevD.66.125013}. This is a purely quantum mechanical symmetry breaking, analogous to what happens for the 2D delta function potential \cite{jackiw1991delta}. We have shown here that a suitable mixed state can be used to restore the broken scaling symmetry. Such a mixed state is associated with a nonvanishing von Neumann entropy. It thus appears that the symmetry which is broken due to the choice of boundary conditions in quantum mechanics may be restored at the expense of a mixed state with nonvanishing entropy. Our analysis provides a connection between the von Neumann entropy and the restoration of scaling symmetry in CQM.

It is known that the near-horizon conformal structure of certain black holes can be described by a CQM, which is associated with a Virasoro algebra \cite{Birmingham2001a,ksg1,Chakrabarti2008}. Using the Cardy formula, the central charge of the Virasoro algebra can be related to the Bekenstein-Hawking formula \cite{carlip}. Here we have argued that the restoration of the symmetry in CQM also leads to an entropy formula. It is thus possible that these two apparently different mechanisms of entropy generation are related, which is under investigation.

\section*{Acknowledgement}
We thank A. P. Balachandran, Bruno Carneiro da Cunha, Manuel Asorey, Jos\'e Garcia Esteve, Fernando Falceto, Filiberto Ares for discussions. NC thanks Prof. Mirjam Cvetic and acknowledges the kind hospitality at the study group of the Departament of Physics and Astronomy at the University of Pennylvania. KG thanks Prof. \'Alvaro Ferraz and acknowledges the kind hospitality at IIP-UFRN, Natal, Brazil, where part of this work was done. ARQ acknowledges the kind hospitality at Departamento de F\'{\i}sica Te\'orica, Facultad de Ciencias, Universidad de Zaragoza. NC is supported by the Programa Ci\^encia sem Fronteiras under CAPES process number 99999.003034/2014-03. SD is supported by CNPq. ARQ is supported by CAPES process number BEX 8713/13-8.

\providecommand{\href}[2]{#2}\begingroup\raggedright\endgroup

\end{document}